\documentclass[%
 reprint,
superscriptaddress,
 amsmath,amssymb,
 aps,
prb,
]{revtex4-2}

\usepackage{natbib}
\usepackage{graphicx}% Include figure files
\usepackage{dcolumn}% Align table columns on decimal point
\usepackage{bm}% bold math
\usepackage{hyperref}
\usepackage{soul}
\begin{document}

%\preprint{APS/123-QED}

%\title{Temperature- and field-dependent point-contact spectroscopy of multiple superconducting gaps in PdTe}
%\title{Field-resolved hierarchy of superconducting energy gaps in PdTe}
\title{Field-resolved hierarchy of superconducting energy gaps in PdTe}

\author{Sangyun Lee}
\email{sangyun.lee@ufl.edu}
\affiliation{Department of Physics and National High Magnetic Field Laboratory High B/T Facility,
University of Florida, Gainesville, Florida 32611, USA}

\author{A. M. Donald}
 %\thanks{These authors contributed equally to this work.}
\affiliation{Department of Physics and National High Magnetic Field Laboratory High B/T Facility,
University of Florida, Gainesville, Florida 32611, USA}

 \author{D. Duong}
 %\thanks{These authors contributed equally to this work.}
  \affiliation{SmartState Center for Experimental Nanoscale Physics, Department of Physics and Astronomy, University of South Carolina, Columbia, South Carolina 29208, USA.}

\author{S. Huang}
\affiliation{SmartState Center for Experimental Nanoscale Physics, Department of Physics and Astronomy, University of South Carolina, Columbia, South Carolina 29208, USA.}

\author{R. Gazizulin}%
\affiliation{Department of Physics and National High Magnetic Field Laboratory High B/T Facility,
University of Florida, Gainesville, Florida 32611, USA}

\author{C. Huan}%
\affiliation{Department of Physics and National High Magnetic Field Laboratory High B/T Facility,
University of Florida, Gainesville, Florida 32611, USA}

%\author{Nicolas Silva}
%\affiliation{National High Magnetic Field Laboratory High B/T Facility and Department of Physics, University of Florida, Gainesville, Florida 32611, USA}

%\author{C. J. Ollmann}%
%\affiliation{National High Magnetic Field Laboratory High B/T Facility and Department of Physics, University of Florida, Gainesville, Florida 32611, USA}

%\author{M. W. Meisel}%
%\affiliation{Department of Physics and National High Magnetic Field Laboratory High B/T Facility,
%University of Florida, Gainesville, Florida 32611, USA}

\author{Rongying Jin}%
\email{rjin@mailbox.sc.edu}
\affiliation{SmartState Center for Experimental Nanoscale Physics, Department of Physics and Astronomy, University of South Carolina, Columbia, South Carolina 29208, USA.}

\date{\today}% It is always \today, today,
             %  but any date may be explicitly specified

\begin{abstract}
Resolving how multiple superconducting energy scales evolve under magnetic field is important for distinguishing multiband, anisotropic, and surface-sensitive superconducting responses. 
Here, we investigate the superconducting state of PdTe using low-temperature soft point-contact spectroscopy in a sealed $^3$He immersion environment. 
The resulting Andreev-reflection spectra reveal two characteristic superconducting energy scales near 0.5 and 1.1 meV that are not adequately described by a single-gap BTK model. 
Both energy scales are progressively suppressed upon warming toward the same superconducting transition temperature $T_c$.
In contrast, their magnetic-field evolution reveals two distinct characteristic fields near 4.5 and 7.5 kG. The lower scale is associated with the loss of the zero-resistance superconducting state, whereas the higher-energy spectroscopic response persists to substantially larger fields. 
Independent field-dependent gap analysis yields suppression scales consistent with those identified from the zero-bias conductance. 
These results reveal a field-resolved hierarchy of superconducting energy scales in PdTe and indicate superconducting responses with distinct magnetic-field robustness.
\end{abstract}

\maketitle

%\tableofcontents

%\section{\label{sec:Introduction}Introduction}

Topological materials, such as topological insulators and superconductors, have attracted extensive interest in condensed matter physics and materials science due to their unique electronic structures and remarkable physical properties~\cite{hasan2010colloquium,qi2011topological,sato2017topological,alicea2012new}. 
In particular, topological superconductors are expected to host Majorana quasiparticles, which obey non-Abelian statistics and provide a platform for fault-tolerant quantum computation~\cite{hasan2010colloquium,qi2011topological,sato2017topological,alicea2012new}. 
Considerable effort has therefore been devoted to realizing topological superconductivity through approaches such as proximity coupling, chemical doping, and heterostructure design~\cite{fu2008superconducting,hor2010superconductivity,sasaki2011topological,wang2012coexistence}. 
However, achieving robust superconductivity with a large volume fraction and well-defined surfaces/interfaces remains challenging in these engineered systems.

This has motivated the search for intrinsic superconductors that naturally host nontrivial topological band structures. 
Among candidate materials, PdTe has emerged as a particularly promising system. 
PdTe is a Dirac semimetal that exhibits superconductivity below $T_c \approx 4.5$~K and hosts nontrivial electronic topology~\cite{karki2012pdte,tiwari2015pdte,chapai2023evidence,yang2023coexistence,duong2026}. 
Bulk superconductivity in PdTe has been well established, and several studies have suggested unconventional features, including multiband superconductivity and possible nodal behavior~\cite{zhao2024multigap,chapai2023evidence,yang2023coexistence}. 

Despite these advances, the superconducting gap structure of PdTe remains under active debate. 
Angle-resolved photoemission spectroscopy (ARPES) has revealed complex surface states associated with its nontrivial band topology~\cite{yang2023coexistence}, raising the possibility of distinct superconducting behavior between bulk and surface. 
On the other hand, thermodynamic and transport measurements have reported signatures consistent with both fully gapped and unconventional superconducting states, leading to seemingly contradictory interpretations of the pairing symmetry~\cite{zhao2024multigap,chapai2023evidence,yang2023coexistence}. 
In particular, specific heat measurements have suggested a multigap superconducting state with a large disparity between gap magnitudes, although the analysis relies on extrapolations to low temperatures and fitting over a limited temperature range, leaving uncertainties in the determination of the smaller gap~\cite{chapai2023evidence}.

While the present study was in progress, an independent point-contact spectroscopy study reported a two-gap superconducting response in PdTe, with conductance spectra from several contacts showing systematic deviations from a single-gap BTK description~\cite{Shi2025}. The reported gap magnitudes and contact-dependent spectral variations differ quantitatively from those found in the present measurements, motivating a direct comparison of alternative gap models followed by temperature- and field-dependent tests of the resulting spectroscopic energy scales.
An important unresolved question is whether these distinct spectroscopic energy scales possess different magnetic-field robustness and how their suppression relates to the loss of global zero resistance. 
The present measurements directly address this issue by tracking the two energy scales deep in the superconducting state under magnetic field.

To determine the nature of the superconducting gap structure, a direct spectroscopic probe is essential. 
Point-contact spectroscopy (PCS), based on Andreev reflection at a normal-metal/superconductor interface, provides a powerful method to directly determine superconducting gap magnitudes and pairing characteristics~\cite{blonder1982transition,daghero2010probing,naidyuk2018surface}. Below the superconducting transition temperature $T_c$, Andreev reflection enhances the conductance through the conversion of electrons into Cooper pairs, providing spectroscopic access to gap-related features.

In this article, we report the investigation of the superconducting state of PdTe using soft point-contact spectroscopy. 
By combining the mechanically stable soft-contact technique with a $^3$He immersion cooling scheme, 
we achieve excellent thermalization and minimize spurious broadening effects at the contact interface. 
Here, we resolve two superconducting energy scales in PdTe and track their evolution with temperature and magnetic field. 
Although both scales are suppressed toward the same $T_c$, their magnetic-field evolution separates into two distinct characteristic field scales. 
The lower scale is associated with the disappearance of the zero-resistance superconducting state, 
whereas the higher-energy spectroscopic response persists to substantially larger fields. 
These measurements reveal a field-resolved hierarchy of superconducting energy scales that cannot be characterized by a single superconducting field scale.

\begin{figure}
\centering
\includegraphics[width=1\columnwidth]{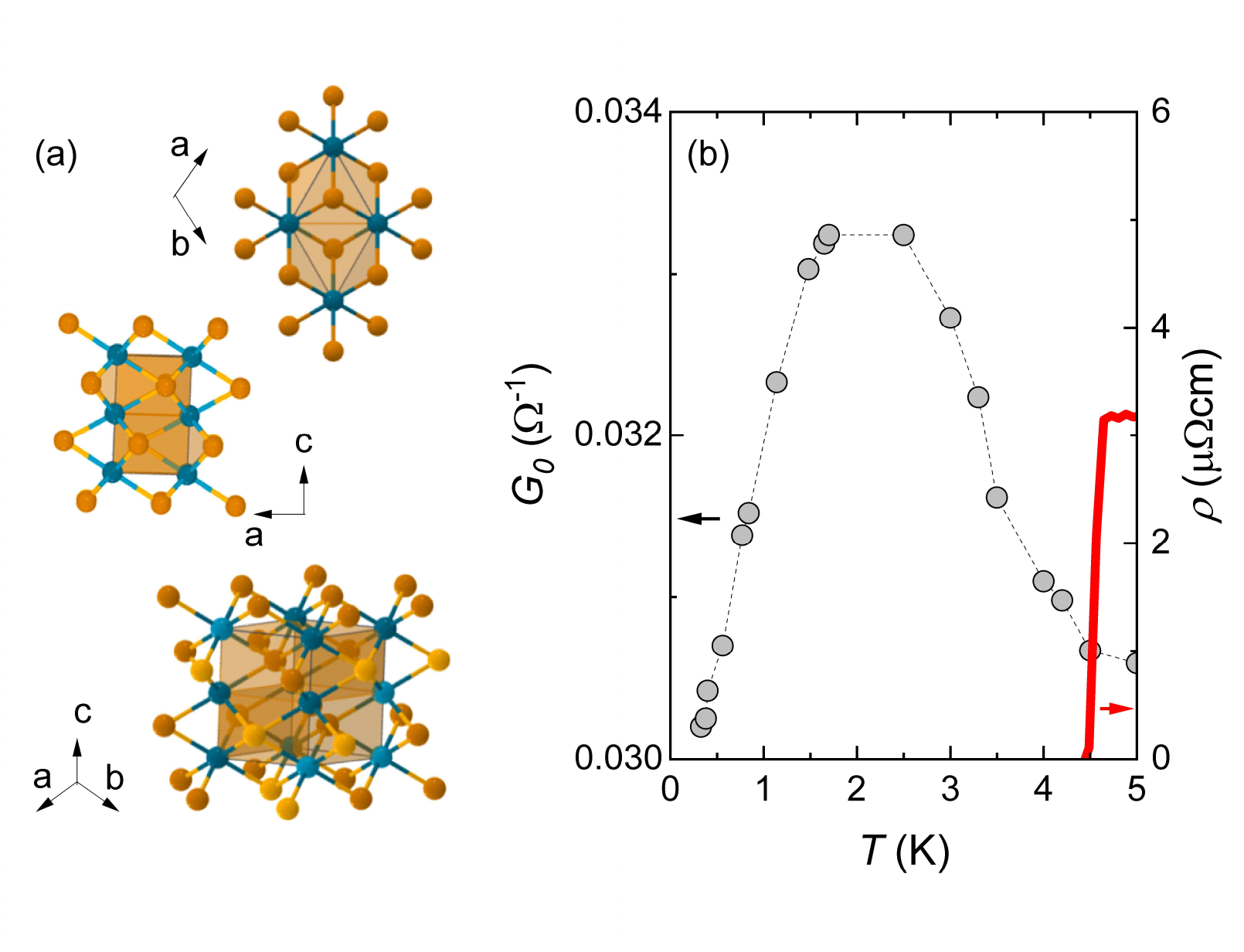}
\caption{\label{fig:rt} (a) Crystal structure of PdTe constructed using the reported lattice parameters and atomic positions from Ref.~\cite{tiwari2015pdte}. 
Blue and brown spheres represent Pd and Te atoms, respectively.
(b) Temperature dependence of the electrical resistance (red line) and zero-bias conductance (open circles) of PdTe.
The resistance and zero-bias conductance show distinct behaviors below $T_c$, indicating that the point contact is not in the thermal regime and preserves spectroscopic sensitivity to superconductivity.}
\end{figure}

\begin{figure}
\centering
\includegraphics[width=1\columnwidth]{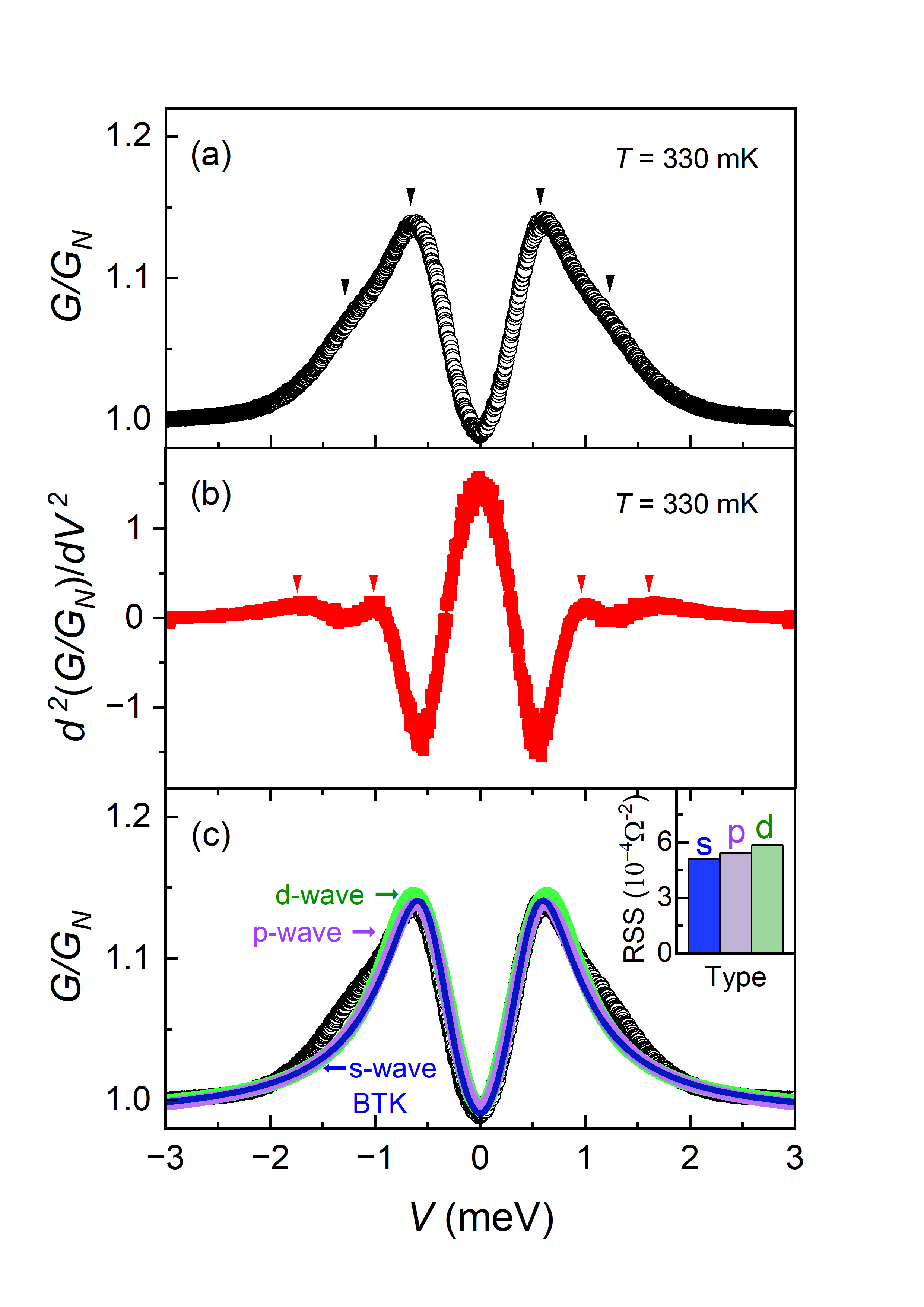}
\caption{\label{fig:fit} (a) Soft point-contact spectroscopy of PdTe. The differential conductance $G$ measured at 330~mK is plotted as a function of bias voltage. Vertical arrows mark additional finite-bias spectroscopic features.
(b) Second derivative, $d^2G/dV^2$, of the differential conductance shown in (a), highlighting two characteristic energy scales.
(c) Normalized differential conductance of PdTe fitted using single-gap $s$-, $p$-, and $d$-wave BTK models. The blue, violet, and green curves represent the least-squares fits obtained using the $s$-, $p$-, and $d$-wave models, respectively. The inset compares the residual sum of squares (RSS) for the three models, showing that the $s$-wave model yields the smallest RSS among the single-gap models considered.}
\end{figure}

\section{\label{sec:res} Spectroscopic energy-scale hierarchy}

Figure~\ref{fig:rt}(a) shows the crystal structure of PdTe, in which Pd and Te atoms form a hexagonal NiAs-type structure~\cite{tiwari2015pdte}. 
The superconducting transition of the measured crystal was first characterized by the electrical resistivity. 
As shown by the red curve in Fig.~\ref{fig:rt}(b), the electrical resistivity decreases sharply and reaches zero at $T_c$ $\approx$~4.5 K, confirming superconductivity in PdTe. 
Soft point-contact Andreev-reflection spectroscopy was performed on PdTe single crystals in a sealed $^3$He immersion environment; further experimental details are provided in Methods.
The zero-bias conductance obtained from the soft point contact is also plotted as a function of temperature in Fig.~\ref{fig:rt}(b). Although both quantities show changes associated with superconductivity, their temperature dependences below $T_c$ are clearly distinct. The resistance drops toward the zero-resistance state, whereas the zero-bias conductance behaves nonmonotonically in the superconducting state, reaching a maximum before decreasing. 
This difference indicates that the point-contact response is not simply governed by the sample resistance or by Joule-heating-induced thermal effects. Instead, the junction retains spectroscopic sensitivity to the superconducting quasiparticle spectrum.

The nonmonotonic temperature dependence of the zero-bias conductance can be caused by multiple reasons. For an ideal normal metal-superconductor junction, the conductance is described by the Blonder--Tinkham--Klapwijk (BTK) model~\cite{daghero2010probing}. For a single superconducting gap, the normalized conductance is expressed as
\begin{equation}
    G_{\mathrm{1gap}}(V)=G_{\mathrm{BTK}}(V,\Delta,Z,\Gamma),
\end{equation}
where $\Delta$ is the superconducting gap, $Z$ is the dimensionless interfacial barrier strength, and $\Gamma$ is the phenomenological broadening parameter. At just below $T_c$, the superconducting gap $\Delta$ begins to open. At a highly transparent interface ($Z$ $\approx$ 0), Andreev reflection enhances the conductance, leading to an increase in the zero-bias conductance. As the temperature decreases further, the superconducting gap becomes larger and the quasiparticle density inside the gap decreases, causing a suppression of the zero-bias conductance. 
To examine whether the low-temperature spectrum can be described within a single-gap framework, we present the soft point-contact spectrum measured at 330~mK in Fig.~\ref{fig:fit}(a). 
The differential conductance, $G=dI/dV$, exhibits a pronounced double-peak structure characteristic of point-contact Andreev-reflection (PCAR) spectra. 
In addition to the main Andreev-reflection peaks, additional finite-bias structures are observed, as marked by the vertical arrows in Fig.~\ref{fig:fit}(a).
To visualize these energy scales more clearly, the second derivative, $d^2G/dV^2$, is shown in Fig.~\ref{fig:fit}(b). The corresponding structures in $d^2G/dV^2$ highlight changes in the spectral curvature and further support the presence of multiple characteristic spectroscopic energy scales.
We compared representative single-gap $s$-, $p$-, and $d$-wave BTK models to the normalized conductance spectrum. 
For the anisotropic single-gap comparisons, the junction orientation was fixed at $\alpha=0$. The $p$-wave and $d$-wave gap functions were taken as $\Delta_p(\theta)=\Delta_0\cos\theta$ and $\Delta_d(\theta)=\Delta_0\cos 2\theta$, respectively, where $\theta$ denotes the quasiparticle injection angle.
As shown in Fig.~\ref{fig:fit}(c), the fully gapped $s$-wave model yields the smallest residual sum of squares (RSS) among the single-gap models considered. 
However, none of the single-gap models reproduces the full spectral line shape, particularly the additional finite-bias structures. These residual features motivate consideration of a superconducting response involving more than one characteristic energy scale.

\begin{figure}
\centering
\includegraphics[width=1\columnwidth]{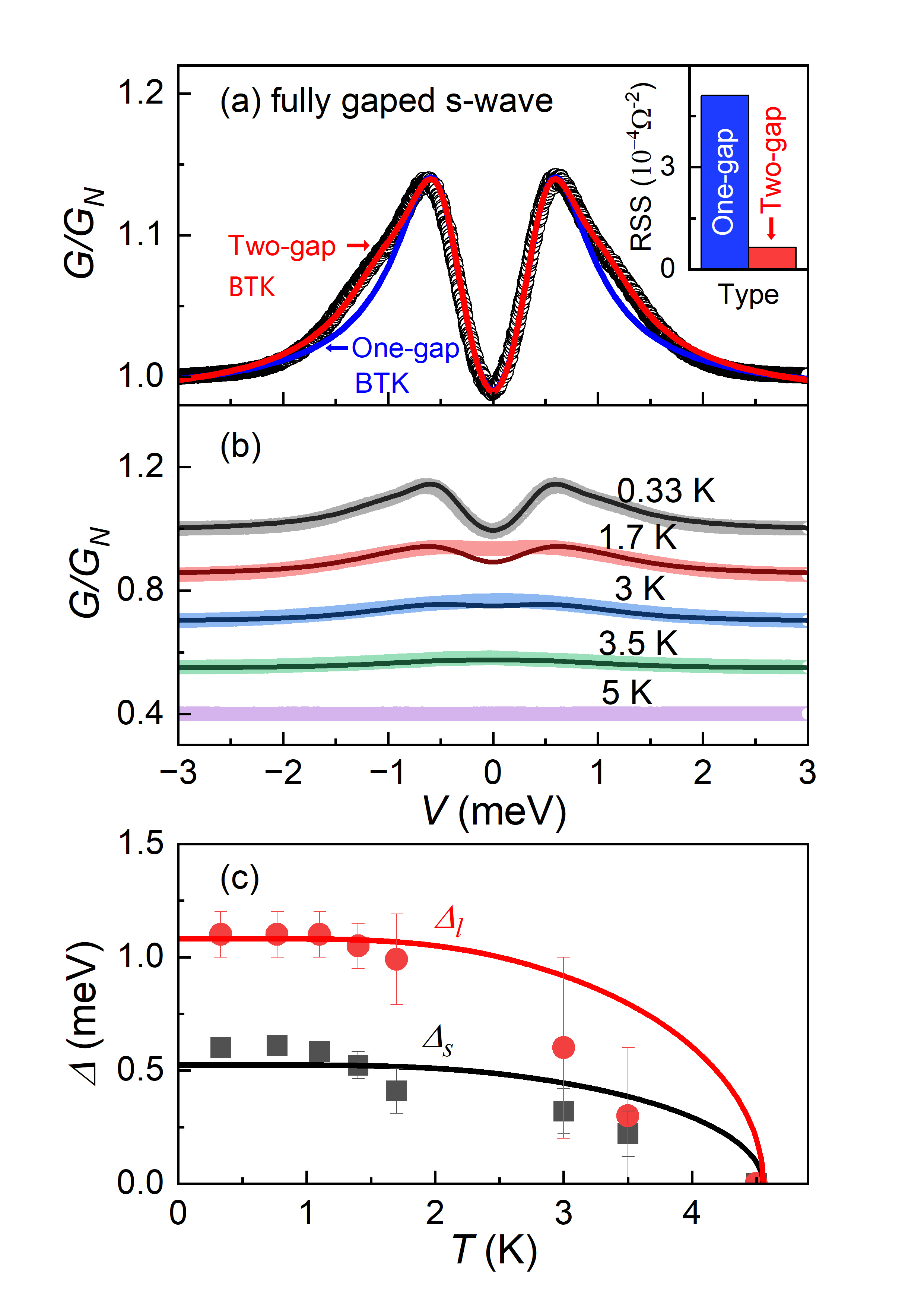}
\caption{\label{fig:temp} 
(a) Normalized differential conductance of PdTe fitted using single-gap and two-gap $s$-wave BTK models. The blue and red curves represent the least-squares fits obtained using the single-gap and two-gap models, respectively. The two-gap fit yields $\Delta_l=1.1$~meV and $\Delta_s=0.5$~meV. The inset compares the residual sum of squares (RSS) for the two models; the RSS of the two-gap fit is approximately eight times smaller than that of the single-gap fit.
(b) Temperature evolution of the soft point-contact Andreev-reflection spectra of PdTe. The solid curves are fits to the two-gap $s$-wave BTK model.
(c) Temperature dependence of the superconducting energy gaps extracted from the two-gap BTK analysis.}
\end{figure}

Having identified the fully gapped $s$-wave model as the best-performing single-gap description, we next tested whether a two-gap $s$-wave model provides a quantitatively improved description of the spectrum. We therefore employed the two-gap BTK model~\cite{daghero2010probing},
\begin{equation}
\begin{aligned}
    G_{\mathrm{2gap}}(V)
    ={}& wG_{\mathrm{BTK}}(V,\Delta_l,Z,\Gamma) \\
    &+ (1-w)G_{\mathrm{BTK}}(V,\Delta_s,Z,\Gamma),
\end{aligned}
\end{equation}
where $\Delta_s$ and $\Delta_l$ are the two superconducting gaps. For a constrained comparison, the two spectral components were modeled using common values of $Z$ and $\Gamma$, while their relative spectral weight was described by $w$.
As shown in Fig.~\ref{fig:temp}(a), the two-gap $s$-wave model provides a substantially improved description of the conductance line shape compared with the single-gap $s$-wave model. 
The RSS of the two-gap fit is approximately eight times smaller than that of the single-gap fit.
The best-fit parameters at $T=0.33$~K are $\Delta_l=1.10\pm0.04$~meV, $\Delta_s=0.5\pm0.1$~meV, $Z=0.51$, $\Gamma=0.28$~meV, and $w=0.1$. 
The small value of $w$ indicates that the smaller-gap contribution dominates the measured conductance spectrum (90$\%$), while the larger gap accounts for the additional finite-bias structure. 
These results suggest that the conductance spectrum of PdTe cannot be described by a single isotropic superconducting gap alone, but is more naturally explained by two superconducting energy gaps.

The superconducting origin of these spectral features is further supported by their temperature evolution. 
Figure~\ref{fig:temp}(b) shows the temperature evolution of the PCAR spectra. 
The solid curves in Fig.~\ref{fig:temp}(b) show two-gap \(s\)-wave BTK fits to the spectra at each temperature. 
The model captures the systematic evolution of the Andreev-reflection features over the measured temperature range.
Upon warming, these structures are gradually suppressed, and the conductance approaches the normal-state background near $T_c$. 
This systematic evolution rules out a temperature-independent junction artifact as the origin of the observed conductance features.
Small deviations between the measured spectra and the two-gap BTK fits at intermediate temperatures may reflect the simplified assumption of two isotropic gap components. 
Additional Fermi-surface sheets, momentum-dependent gap anisotropy, and temperature-dependent spectral broadening or weighting may contribute to the detailed conductance
line shape.

The superconducting energy gaps extracted from the two-gap BTK analysis are summarized in Fig.~\ref{fig:temp}(c). 
Both extracted values decrease with increasing temperature and are suppressed when approaching $T_c$. The solid lines in Fig.~\ref{fig:temp}(c) represent the BCS-like temperature dependence,
\begin{equation}
\Delta_i(T)=\Delta_i(0)\tanh\left[1.74\sqrt{\frac{T_c}{T}-1}\right],
\end{equation}
and are included as guides to the eye. 
Their simultaneous temperature-dependent suppression supports the superconducting origin of the two spectroscopic energy gaps. 
The fact that both energy scales are suppressed toward the same $T_c$ further suggests that they are associated with the same superconducting transition rather than with independent extrinsic phases.

A comparison between the present spectroscopic energy gaps and those inferred from bulk specific-heat measurements is instructive. 
A two-gap analysis of the electronic specific heat yielded gap values of approximately 0.37~meV and 1.93~meV~\cite{chapai2023evidence}. 
The smaller gap is reasonably close to the present value of approximately 0.5~meV, whereas the higher-energy gap resolved here, approximately 1.1~meV, lies below the large gap inferred from specific heat. These differences may reflect the distinct weighting of Fermi-surface sheets and momentum-space regions by thermodynamic and point-contact probes. The reported values may therefore represent different projections of the same underlying superconducting gap structure rather than mutually incompatible superconducting states.
A recent PCS study also reported a two-gap response in PdTe, although with smaller characteristic gap energies~\cite{Shi2025}. The two contacts examined in the present work yield a larger scale near 1.1~meV and a smaller scale near 0.5--0.6~meV. Such differences may reflect contact-dependent weighting of different Fermi-surface sheets or momentum-space regions.

\begin{figure}
\centering
\includegraphics[width=1\columnwidth]{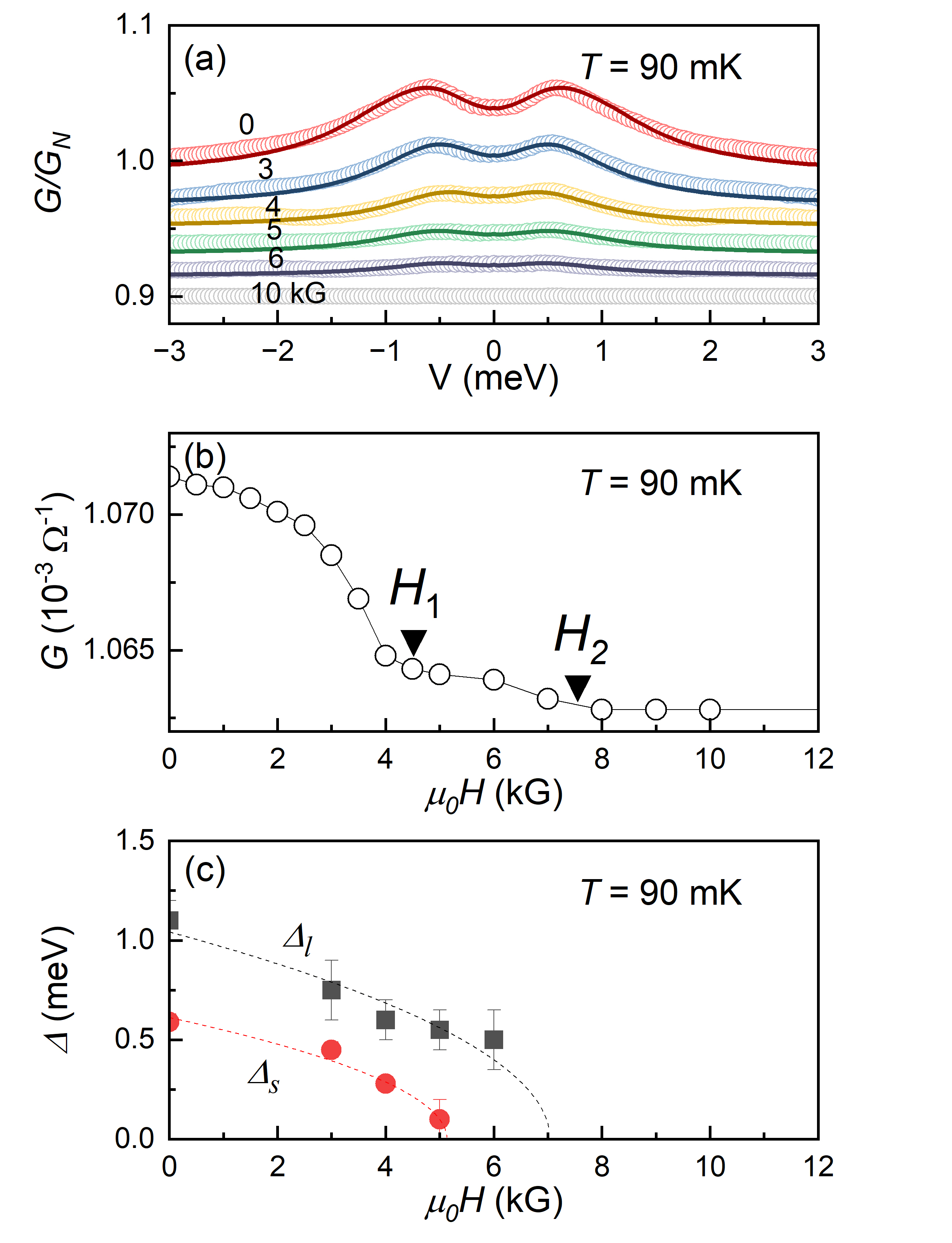}
\caption{\label{fig:field} (a) The magnetic-field evolution of the differential conductance spectra of PdTe measured at 90~mK for fields of 0, 3, 4, 5, 6, and 10~kG.
(b) Magnetic-field dependence of the zero-bias conductance of PdTe. 
Vertical arrows mark characteristic fields, denoted $H_1\approx4.5$~kG and $H_2\approx$7.5~kG, associated with the suppression of the two spectroscopic energy gaps.
(c) Magnetic-field dependence of the superconducting energy gaps extracted from the two-gap BTK analysis.}
\end{figure}

The magnetic-field dependence of the PCAR response is examined next. Figure~\ref{fig:field}(a) shows the field evolution of the full differential conductance spectra at $T$ = 90 mK. At zero field, the spectrum exhibits clear Andreev-reflection features. As the field increases, these features are gradually weakened and broadened, rather than vanishing abruptly. This gradual suppression indicates that the spectroscopic superconducting response evolves over more than a single characteristic field scale. This behavior is more clearly reflected in the magnetic-field dependence of the zero-bias conductance as plotted in Figure~\ref{fig:field}(b). With increasing field, the zero-bias conductance evolves toward the normal-state value and exhibits a slope-change region that terminates near 4.5~kG, together with a second characteristic feature near 7.5~kG, as indicated by the vertical arrows. 
The lower characteristic field near 4.5~kG is of the same order as the zero-resistance transport scale reported previously in transport measurements. 
In particular, the resistive transition reported in Ref.~\cite{zhao2024multigap} extends to a similar field scale when the zero-resistance boundary, rather than the onset criterion, is considered. The higher characteristic field near 7.5~kG, however, extends beyond this transport scale.

To confirm the above observation, the field-dependent measurements were performed using a different spectroscopic contact from that used for the temperature-dependent measurements. At zero field and 90~mK, this contact yields gap values of approximately $\Delta_l=1.1$~meV and $\Delta_s=0.59$~meV, close to those obtained from another junction. Thus, although the absolute gap values vary slightly between junctions, both data sets consistently exhibit two characteristic superconducting energy gaps. In the field-dependent BTK analysis, $Z$, $\Gamma$, and the spectral weighting factor $w$ were fixed to the zero-field values obtained for the same junction, $Z=0.36$, $\Gamma=0.47$~meV, and $w=0.2$, respectively, while only $\Delta_s$ and $\Delta_l$ were allowed to vary. 
The constrained fitting procedure was adopted to reduce parameter covariance and to isolate the magnetic-field evolution of the two energy scales.

The magnetic-field dependence of the two superconducting energy gaps is shown in Fig.~\ref{fig:field}(c). To characterize the field dependence of the extracted gaps, each energy scale was fitted using the phenomenological form
\begin{equation}
\Delta_i(H)=\Delta_{i,0}\sqrt{1-\frac{H}{H_{i}}},
\end{equation}
where $i=s,l$, $\Delta_{i,0}$ is the zero-field gap amplitude, and $H_{i}$ is the characteristic field at which the corresponding gap is suppressed. 
The phenomenological fits indicate that $\Delta_s$ and $\Delta_l$ vanish approximately at 5.1$\pm$0.2~kG and 7$\pm$0.5~kG, respectively. These values are in good agreement with the two characteristic field scales of approximately 4.5 and 7.5~kG independently identified from the zero-bias conductance. 
This correspondence suggests that suppression of the smaller energy scale is closely associated with the loss of the zero-resistance superconducting state, whereas the larger energy scale remains detectable to a higher characteristic field.

The central result is not only the presence of multiple spectroscopic energy scales, but their contrasting magnetic-field robustness. 
Although both scales are suppressed toward the same $T_c$, their field evolution separates into two characteristic suppression scales near 4.5 and 7.5 kG.
This establishes a field-resolved hierarchy of superconducting responses that is not captured by a single superconducting energy or field scale.
Among the single-gap models considered, the fully gapped $s$-wave model yields the lowest residual error, while a two-gap $s$-wave model provides a substantially improved description of the full conductance line shape. 
The two extracted energy gaps are suppressed upon warming toward $T_c$ and exhibit distinct characteristic suppression fields near 4.5 and 7.5~kG.

This interpretation is compatible with the multiband electronic structure of PdTe~\cite{chapai2023evidence}. 
Previous studies have discussed PdTe as a spin-orbit-coupled topological semimetal candidate with multiple Fermi-surface sheets, and thermodynamic and transport measurements have suggested deviations from a simple single-gap BCS description~\cite{zhao2024multigap,chapai2023evidence}. 
In this context, the present soft point-contact spectroscopy provides an energy-resolved probe of the superconducting state and shows that the conductance spectra are more naturally described by two superconducting energy gaps.

Although the single-gap $p$- and $d$-wave models considered here yield larger residuals than the fully gapped $s$-wave model, this comparison alone does not uniquely determine the microscopic pairing symmetry. The present data also do not uniquely distinguish among two fully opened gaps, a strongly anisotropic gap with multiple characteristic extrema, and a combination of bulk- and surface-sensitive superconducting responses. 
The larger spectroscopic energy scales obtained from PCAR compared with those inferred from earlier bulk magnetic and thermodynamic measurements~\cite{karki2012pdte} may be consistent with a surface-sensitive contribution to the measured spectra. 
If surface superconductivity is indeed nodeless and more field robust than the bulk superconducting state, PCAR may retain superconducting spectral features to fields beyond the bulk transport scale. In several topological semimetals, enhanced and field-robust surface superconductivity has been reported and associated with nontrivial surface electronic states~\cite{Schimmel2024,Moreno2026,Mascot2022}. For PdTe, recent work has also indicated a correlation between the superconducting transition temperature and electronic density-of-states-related quantities~\cite{duong2026}. Further microscopic studies of the surface electronic structure and surface density of states will be needed to determine whether the higher-energy and higher-field spectroscopic features observed here are associated with surface superconductivity.

In summary, low-temperature point-contact spectroscopy resolves two characteristic superconducting energy scales in PdTe, both of which are
suppressed toward the same superconducting transition temperature.
Their magnetic-field evolution, however, is markedly different, with characteristic scales near 4.5 and 7.5 kG. 
The lower field scale is closely associated with the loss of zero resistance, whereas the higher-energy spectroscopic response persists to substantially larger fields. 
The distinct temperature and field evolution of the two spectroscopic components therefore reveals a hierarchy of superconducting energy scales that cannot be described by a single superconducting field scale. 
This behaviour is compatible with a multiband superconducting state and may also reflect anisotropic or surface-sensitive contributions to the measured response. 
These results provide a spectroscopic framework for disentangling superconducting components with different magnetic-field robustness in PdTe.

\section{\label{Methods}Methods}

Soft point-contact spectroscopy (Soft PCS) is a sensitive probe of superconducting gap structures and interfacial electronic properties~\cite{blonder1982transition,daghero2010probing,naidyuk2018surface}. In this work, Soft PCS measurements were performed on PdTe single crystals to investigate their superconducting gap structure. Electrical contacts were formed using a soft-contact technique, in which small silver grains were gently deposited onto the sample surface~\cite{blonder1982transition,daghero2010probing}. 
This approach minimizes mechanical strain and preserves the intrinsic electronic properties of the material, while ensuring stable and reproducible contacts.

After systematic screening of multiple crystals and contact configurations, stable junctions exhibiting spectroscopic characteristics were selected for detailed measurements. Measurements were performed using a sealed $^3$He immersion cell mounted on and thermally anchored to the mixing chamber of a dry dilution refrigerator system (Bluefors)~\cite{donald2026sub10mk}. 
Both the sample and the electrical leads were fully immersed in liquid $^3$He, providing efficient thermalization and reducing local heating effects at the point-contact interface~\cite{donald2026sub10mk}.

\section*{Acknowledgements} \label{sec:acknowledgements}
    This work was substantially revised following its presentation at the VI International Conference "Condensed Matter $\&$ Low-Temperature Physics".
    We thank M. W. Meisel and Y. G. Naidyuk for insightful discussions that helped improve the manuscript.
    This work was performed at the National High Magnetic Field Laboratory High $B/T$ Facility on the campus of the University of Florida. 
    The National High Magnetic Field Laboratory is supported by the National Science Foundation through Grants DMR-2128556 and the State of Florida. 
    This work was partially supported by the National High Magnetic Field Laboratory through the NHMFL User Collaboration Grants Program (UCGP). Work at USC (D.D., S.H., and R.J.) was supported by the National Science Foundation through Grants DMR-2532355. 

\bibliography{Main_v58}% Produces the bibliography via BibTeX.

\end{document}